# Revisiting elliptical satellite orbits to enhance the *O3b* constellation

Lloyd Wood
University of Surrey alumni
Sydney, Australia.
L.Wood@society.surrey.ac.uk

Yuxuan Lou
University of Surrey alumni
London, England.
dennis.lou@forever.surrey.ac.uk

Opeoluwa Olusola
University of Surrey alumni
Guildford, England.
mr.ope.olusola@ieee.org

**Abstract** – Highly elliptical orbits can be used to provide targeted satellite coverage of locations at high latitudes. We review the history of use of these orbits for communication. How elliptical orbits can be used for broadband communication is outlined. We propose an addition of known elliptical orbits to the new equatorial *O3b* satellite constellation, extending *O3b* to cover high latitudes and the Earth's poles. We simulate the *O3b* constellation and compare this to recent measurement of the first real Internet traffic across the newly deployed *O3b* network.

1. INTRODUCTION

How useful are elliptical orbits for communication? How are they used? Early low-orbiting satellites were launched into Highly Elliptical Orbits (HEO) as a result of not having much control over trajectory. Circular orbits with minimal eccentricity offer consistent altitudes, with the benefits of consistent free space losses and link budgets throughout the orbit, and soon became the norm. Highly elliptical orbits fell from favour for communications use.

Most communication through orbiting satellites is now via the circular geostationary ring, at 35,786 km above the Earth's equator. There, the angular velocity of the satellites' motion matches the eastward rotation of the Earth's surface. The result is that the satellites appear to be approximately fixed in the sky, modulo minor perturbations. Satellite dish reflectors can be fixed in position on the ground to point at each satellite in its allocated orbital slot. Satellites in this orbit have a wide view of the Earth, seeing a belt between approximately 75° of latitude North and South. At higher latitudes, the geostationary ring lies below the local horizon, and these satellites are out of view.

The popularity of geostationary satellites means that the most common elliptical orbit used is actually the Geostationary Transfer Orbit (GTO), an interim Hohmann transfer orbit that places the satellite in Geostationary Earth Orbit (GEO) at minimum cost.

For coverage at higher latitudes, where the geostationary arc is blocked by terrain, so-called quasi-geostationary elliptical orbits can be used. These require a number of satellites in similar, complementary orbits. While each satellite passes around apogee (the highest point), its angular velocity matches the rotation of the Earth's surface while the satellite moves away from or toward the Earth. The inclination of these orbits to the Equator, at 63.4°, prevents precession and permits this matching of angular velocities. As each satellite leaves apogee, another satellite entering apogee takes its place, with handover of communication, from the satellite falling to perigee (closest approach) to the satellite rising to apogee. This enables targeted coverage of selected high-latitude locations, using a minimal number of satellites.

2. MOLNYA AND TUNDRA ELLIPTICAL ORBITS

Molniya and Tundra orbits were exploited by the Soviet Union, for television broadcast to high-latitude cities [1]. Geostationary satellites are low in the Russian sky, hence their name Gorizont (Russian for 'horizon').

A Molniya (Russian for 'lightning') orbit is shown alongside other orbits in Fig. 1. This needs a constellation of three satellites in 12-hour orbits to provide continuous coverage of a selected longitude from an apogee point just beyond geostationary orbit, with eight hours of use per satellite. Similar overlapping coverage is also provided to a point in the same hemisphere, 180 degrees opposed, so Russian Molniya systems could also provide good coverage of continental North America.

Tundra orbits, whose apogee lies at almost twice the altitude of geostationary orbit, rely on two satellites in 24-hour highly elliptical orbits to provide continuous coverage of the chosen area, with 12 hours of use per satellite between handovers.

Given the good view of the poles enabled by these constellations, Canada has also proposed Molniya use for several systems, including POLARSAT for meteorological observation [2]. Sirius Radio deployed satellites in modified Tundra orbits in a three-satellite Molniya-like constellation [3]; the large propagation delay does not matter for the radio broadcast application. Satellites are uniquely suited for broadcast to large areas, and television and radio can exploit this.

The choice of orbital period relative to the Earth's sidereal day helps ensure that the apogee remains fixed from the Earth's point of view. As each apogee point appears fixed in space from the ground, even though the satellites are moving up or down, fixed dishes on the ground can be used for communication through these satellites.

3. THEORETICAL ELLIPTICAL ORBITS

Highly elliptical orbits have been explored, designed, and promoted extensively by John E. Draim and colleagues [4].

*3.1 Global coverage with minimum number of satellites*

Satellites are expensive to build, launch, and operate. The resulting rule of satellite constellation system design is that, however many satellites you plan, it's too many. The number of satellites must be minimized. Many proposed systems have been redesigned to use fewer satellites to reduce costs.

Arthur C. Clarke recognized that a minimum of three geostationary satellites in the known circular stationary orbit, now sometimes called the Clarke orbit, could span most of the Earth, bar the polar regions [5]. Clarke famously did not patent the idea, later remarking that a patent was just a license to be sued [6]. Inmarsat's Global Xpress system is a recent example of this classic three-satellite Clarke constellation, though without the intersatellite links or educational television applications that Clarke envisaged.

Draim recognized that four satellites in large elliptical orbits, forming the moving vertices of a tetrahedron, could provide minimum continuous coverage of the entire Earth, if the Earth was always inside the tetrahedron [7]. He then patented his idea [8]. This elliptical constellation is unusual, in that it requires continuous transmission for communication from each satellite throughout its entire orbital period. This includes transmission through the perigee at 42,950 km altitude, and not just from near the high apogee at 78,160 km altitude, to achieve full global coverage. With orbital inclinations of 31.3° to the Equator, the satellites' movement is not matched to the Earth's rotation.

This tetrahedral system's orbital eccentricity $e$ of 0.263 is relatively low compared to Molniya at 0.7 or Tundra at 0.4, so, strictly speaking, this is not *highly* elliptical. (This has not been implemented. It would need visibility at very low elevation angles, multiple tracking antennas for smooth signal handover, and very radiation-resistant electronics on the satellites.)

We have made our simulations of the Moniya, Tundra and Draim constellations, and many other systems, available for use with our *SaVi* satellite constellation simulator [9].

*3.2 Further elliptical work*

Draim has, over decades, provided a wide variety of other explorations of the possibilities of elliptical orbits, including 'COBRA' [10] and 'Droplet' [11], and many of his ideas exploring use of highly elliptical orbits have also appeared in the patent literature. Patent applications and US Federal Communications Commission (FCC) frequency filings often provide more technical information on the design of planned satellite constellation systems than is found in other literature, and are worth examining for that reason.

In the academic literature, flower constellations promise to bring a coherent underlying mathematical theory for repeating-groundtrack elliptical systems to wider attention [12].

4. SATELLITE CONSTELLATIONS IN PRACTICE

*4.1 Optimism*

Elliptical orbits have featured much less in the satellite communication systems that have been constructed. It is worth overviewing the history of these systems to understand why.

In the 1980s and '90s, terrestrial mobile cellular telephony, particularly the European GSM standard, had not yet achieved the ubiquity it has today. In the United States, competing analogue cellular systems provided sporadic coverage in cities but not between them. This provided the impetus to provide telephony via satellite communication, and supported the interest in and financing of commercial systems. All of these systems intended communication with portable handsets, providing voice telephony, messaging, and low-rate (2400/9600bps) data services.

Use of handsets with L-band omnidirectional antennas, within the limits imposed by batteries and acceptable radiation exposure to human tissue, dictated link budgets within tight margins. These required the use of orbits much closer to Earth than the geostationary orbit, to minimize free space losses. These lower orbits in turn demanded a large number of satellites in each low-orbit constellation to ensure continuous overlapping coverage as the satellites move over the Earth's surface. The large number of satellites increases construction, launch and operation costs, and the continuous movement of satellites that is required prevents gradual incremental deployment of service to regions to slowly scale up a service to global coverage.

With one notable exception, *Ellipso*, all of these voice telephony systems relied on circular orbits.

*4.2 Aftermath*

Many companies engineering satellite telephony systems entered bankruptcy proceedings to write off their high development costs, as the expected markets failed to materialize and everyone began using cellphones instead. Even so, some of these systems were built:

a) *Iridium* [13], based on a circular Walker star constellation geometry [14]. The original design of *Iridium* planned 77 active satellites at 780 km altitude in Low Earth Orbit (LEO), and was named for the element of the periodic table with 77 electrons. A redesign to 66 active satellites to reduce costs did not prompt a renaming to the element *Dysprosium*, or Latin for 'bad approach' [15]. *Iridium* uses a mesh of Ka-band intersatellite links to interconnect its satellites, which have onboard switching. This enabled it to scale down terrestrial operations and costs when users did not materialize, and to support its entire constellation and worldwide traffic from very few gateway stations.

b) *Globalstar* [16], using a Ballard rosette [17] rather than a Walker star geometry, at 1400 km altitude in LEO. Unlike *Iridium*, *Globalstar* did not implement inter-satellite links to interconnect its satellites, and so can only complete communications in its small satellite footprint areas near land, where terrestrial gateway stations could be constructed and maintained – and its inclined orbits of the rosette do not pass near the poles. So, despite having to maintain an extensive and expensive terrestrial network to support its space segment, *Globalstar* has never offered global service, and was never a Walker star.

Despite their naming handicaps, these services launched and became operational – and filed for bankruptcy protection to write off development costs as the expected markets did not materialize and their anticipated users satisfied their communication needs with the use of a cellphone and the cheapest minimum of zero satellites [18]. Both Iridium and Globalstar are still operational today as companies and as services, and are deploying their second-generation constellations with broadband Internet communication in mind. However, the bankruptcy filings that allowed these systems to survive led to stillborn competitors in this space as investment ceased.

Some other efforts (TRW's *Odyssey*, Inmarsat's *ICO* and Ellipsat's *Ellipso*) eventually merged their forces into *ICO* and are now but a footnote to history. The circular MEO Medium Earth Orbits (not 'Middle,' even when each orbit is a ring) used in *ICO* and *Odyssey* were, at around 10,400 km altitude, lower than the altitudes at 20,000 km and above that are used by the MEO *GPS, Glonass* and planned *Galileo* navigation systems. A single satellite of the *ICO* constellation was launched successfully to 'Intermediate Circular Orbit,' or MEO, in June 2001. It and its unlaunched sister satellites eventually became a tax writeoff in 2012 [19].

With the growth of terrestrial cellular telephony for business travellers, satellite telephony remains a niche market for remote sites, explorers, rig workers, emergency communications, military operations, and the like.

5. *ELLIPSO* FOR VOICE TELEPHONY

Unlike other LEO and MEO satellite constellation systems intending to use circular orbits to provide voice services, the planned *Ellipso* constellation was unusual in requiring satellites both sharing a circular MEO ring, and using elliptical orbits with an apogee height matching that of the altitude of its equatorial MEO ring. Its orbital geometry was designed and patented by Draim [20].

The six satellites in *Ellipso*'s equatorial ring at 8,040 km altitude (the so-called *Concordia* ring) were intended to provide a ring of coverage around the earth between 55° of latitude North and South. This ring was to be complemented by a constellation of ten satellites, five in each of two highly elliptical orbits, inclined at 63.4°. These orbits have an eccentricity $e$ of 0.347 and an apogee of 7,846 km altitude, giving a low perigee of a mere 520 km altitude. This was the *Borealis* subconstellation, giving complete coverage of the northern hemisphere as its satellites followed one another and their apogees moved around the Earth. *Ellipso* coverage from these two subconstellations is shown without individual spotbeam patterns, but with simple limits of satellite visibility and possible signal coverage for a minimum ground terminal elevation angle of 25°, in Fig. 2.

The southern hemisphere at greater latitudes than covered by the *Concordia* MEO ring was simply dismissed as too small a market; David Castiel, the then-head of Ellipsat, remarked that 'my business plan can do without the people on Easter Island' [15].

6. THE PUSH TO BROADBAND CONSTELLATIONS

*6.1 Optimism*

The rapid growth of the Internet led to more demand for broadband data services and Internet access. Many broadband LEO constellations were proposed in the 1990s.

Craig McCaw's *Teledesic* proposal [21], which later merged with Motorola's *Celestri* [22] and reduced its planned number of satellites a number of times while increasing its planned altitude, was a leading contender, as was Alcatel's *Skybridge* [23].

Elliptical orbits were again proposed for broadband use, once again thanks to Castiel and Draim, with *Virtual GEO* [24] using Draim's Virgo orbits [25][26]. Denali Telecom's *Pentriad* proposed reusing Molniya orbits [27]. These and many other proposals existed primarily as legal paperwork, to take advantage of US Federal Communications Commission (FCC) allocations of high frequencies at Ka-band and at V-band.

### 6.2 Aftermath

These planned broadband proposals were not built. The bankruptcy filings of the voice systems and the difficulty of their construction and launch had introduced a note of caution. Doubts about the eventual market, along with previous lack of success with earlier FCC frequency allocations and the bursting of the pre-2000 internet technology economic bubble, rather cooled enthusiasms.

Eventually, and entirely in line with Clarke's dismal outlook on patents, Draim sued Castiel's organisations over patent bounties [28]. A number of expiring patents, related to Ellipso and Virtual GEO, later came up for auction in 2012 [29][30]. Their current ownership is unclear.

As with cellular telephony, cheaper terrestrial means of achieving the same result, without satellite use, became widespread. With the notable exception of satellite television broadcast, satellite use for communication remains a niche for those in remote areas who need it. Geostationary satellites continue to provide data to fixed dishes, though rarely at rates considered broadband by terrestrial users. Widespread broadband satellite use has yet to be realised, although this is now being attempted by O3b Networks.

## 7. DEVELOPMENT OF *O3b*

### 7.1 Design

This 'other three billion' satellite constellation is named for the large population in the equatorial regions, particularly Africa and Latin America, lacking broadband access. The *O3b* satellite constellation is intended to provide satellite coverage and service to these people [31]. O3b Networks aims to provide broadband trunking connectivity to local internet and voice/mobile telephony providers, using satellites in a single MEO equatorial ring at 8,062 km altitude, with a period of 288 minutes. This altitude, lower than geostationary orbit, gives a shorter light travel time and lower propagation delay than the half-second round-trip time of geostationary satellites [32]. This leads to decreased communication latency and faster network communications. Like *Globalstar*, *O3b* does not use intersatellite links, but the wider coverage from its MEO altitude leads to larger satellite footprints, requiring fewer groundstations to complete signal paths. If successful in its aims, O3b Networks will be the first to offer non-geostationary broadband communication to a large amount of the world's population.

Non-geostationary equatorial rings have been proposed previously. In the 1990s, Orbital Sciences Corporation proposed *Orblink*, with seven satellites at 9,000 km altitude [33]. The University of Surrey suggested *LEqO*, with eight satellites orbiting at just below 1,000 km altitude [34].

We used the network simulator *ns*-2 to simulate the *O3b* constellation, after adding simulation of elliptical orbits to *ns*-2 [35]. This enhances *ns*-2's existing support for simple circular orbits for simulating popular satellite constellation designs [36][37].

A sample simulated communication, with a mean 150 ms one-way delay between ground terminals, is shown in Fig. 3. This illustrates the satellite passes, with delays decreasing as each satellite passes nearest to local zenith, and handover to the following satellite at the peak of maximum delay. These curves can be contrasted with the delay curves for other simulated constellations presented in [38].

*7.2 Construction*

The first four of O3b Networks' satellites were launched in June 2013, and were undergoing testing and beginning operational use as this was written. Coverage for testing is currently across the Tropics, with test locations including the Cook Islands [39]. The Cook Islands site in Rarotonga lies near the limits of current continuous overlapping coverage from four satellites. This is shown in Fig. 4. This service transitioned from testing to commercial service on 15 March 2014 [40].

On 10 January 2014 we were able to ping a device in the Cook Islands continuously across the public Internet, once a second over a twenty-four hour period, from computers located in Surrey, England, in the United Kingdom, and from California in the United States. Internet path delays from those computers to the Cook Islands via the Hawaii gateway over the course of a day are given in Fig. 5.

With geostationary satellites, visible satellite wander over the course of a day is small, leading to fixed delays. For LEO satellites, there is pronounced motion over the Earth during passes, leading to rapid variation in communication delay – but this variation is of the order of 10 ms or so, making satellite movement unclear from a ping with typical granularity of 4 ms. The motion of O3b MEO satellites leads to delay variation in a range that can be clearly instrumented and shown by simple pings. These first results show the movement of *O3b* satellites and handover between orbital passes as a lower bound on the varying ping delays across the Internet between our computers and the Cook Islands. This lower bound matches the shape of the delay curves generated by our simulation.

There are always minimum path and processing delays across the Internet to be added to the satellite propagation delay. These added delays vary with location, so the overall delay to the Cook Islands via the Internet and *O3b* from the UK is larger than that from the US. Internet paths and their resulting delays may be decreased by better peering with other networks, giving improved routing and shorter paths. Even so, the overall mean path delay remains less than that across a single geostationary satellite link.

O3b Networks aims to add more satellites to its orbital ring over time, to increase system capacity as well as to increase the width of the belt between which continuous overlapping coverage can be provided. Four further satellites are planned for launch in 2014. A planned constellation with eight operational satellites is shown in Fig. 6. The aftermath of the effects that O3b Networks will have on satellite communication are not yet clear.

*7.3 Extending coverage?*

There are, however, at least another four billion people on Earth, many living in the Northern hemisphere outside the equatorial belt covered by *O3b*. The similarity of *O3b*'s MEO orbit and the *Ellipso Concordia* orbit – almost identical equatorial altitudes – is striking. If eight satellites can provide coverage and service to three billion people, does adding another ten to cover another four billion people make sense? Would adding a *Borealis*-like elliptical component to *O3b* to cover the rest of the northern hemisphere be possible?

# 8. ADJUSTMENTS NEEDED TO ADD TO *O3B*

## 8.1 On the ground

*O3b* relies on high frequencies at Ka-band, with power control and adjustment against rain fade, to achieve its high data rates to sites. The eastward motion of its MEO satellites always requires tracking, rather than fixed, reflectors at ground stations. These dishes follow the satellites as they move around the circular orbit. Tracking antennas are also necessary for ships relying on geostationary satellites, so O3b Networks is not disadvantaged in the maritime market.

For continuous communication, two dishes can be used, so that one can slew back and acquire the next satellite as the other is in use, with smooth handover without a brief interruption. A third dish can be installed as a ready spare to the active two. This redundant setup reflects O3b Networks' emphasis on broadband backhaul for infrastructure, rather than satellite telephony's emphasis on personal use for low-rate voice calls.

As the communication altitudes and link budgets for O3b's MEO satellites and for the *Borealis* elliptical orbits are much the same, reusing this tracking capability to track the moving apogees of *Borealis* elliptical orbits would merely be a matter of reprogramming the pointing of the dishes. Dishes would track satellites either in the geostationary MEO arc, or the 63.4° MEO apogee arcs in the northern (and perhaps even southern) hemisphere – but would follow satellites in only one of these arcs.

The equatorial ring, northern ellipses and southern ellipses would form three separate subconstellations. The satellites act as sophisticated bent-pipe repeaters, which conduct handoffs to other satellites within the same circular or elliptical subconstellation. More gateway stations would be needed at high latitudes to complete connections through apogee.

## 8.2 In space

However, more work would be needed on the space segment. Added coverage from the elliptical northern hemisphere subconstellation is shown in Fig. 7. The complete constellation, with the addition of the Borealis orbits inclined at 63.4°, with eccentricity $e$ of 0.347, apogee of 7,846 km altitude and perigee of 520 km altitude, is shown in Fig. 8. O3b Networks already has operational satellites engineered for the MEO circular orbit, but reusing the base satellite design for elliptical orbits would need:

   a) Consideration of the varying radiation environments experienced by the satellites, as they pass to and from perigee, transiting the inner Van Allen radiation belt, and the effects of this on the lifetimes and reliability of the satellite components.
   b) Control of the satellite transponders, either disabling transmission or significantly lowering the power when the satellites are not at apogee. Transmission would be disabled below 6,000 km altitude.
   c) Alteration to tracking algorithms for targeting of ground dishes during the orbit and for ensuring smooth handover.
   d) Changes to the Doppler buffers to compensate for satellites receding and approaching as they pass through apogee.
   e) Changes to solar arrays, sun tracking and pointing, and battery cycling.

There are no doubt other re-engineering issues to address in adopting the base satellite design for use with elliptical orbits.

*8.3 In the market*

This addition of an already well-thought-out and documented elliptical proposal to the circular *O3b* design is certainly technically possible, once any lingering frequency rights or legal patent rights issues are resolved.

Far more pressing questions are whether the business case for doing this technical work would be justifiable, and whether there is enough of a market in the wilds of Siberia or in the depths of Canada to justify the engineering investment required.

Unlike the circular component, the elliptical component is not as simply scalable, as satellite coverage overlap and handover become more complex. The northern elliptical sub-constellation would cover many more people, but many of those people have many other choices for network connectivity. It may only be addressing a smaller additional base of potential customers, with diminishing returns.

## 9. CONCLUSIONS

A review of planned uses of elliptical orbits has found a previous proposal that complements the orbital geometry of a broadband satellite system currently under construction.

Adding satellites in elliptical orbits to *O3b* can be technically feasible with considerable reuse of existing technology. There may be demand and a business case for broadband Internet services to remote high-latitude locations that can best be served by satellite, and O3b Networks' equipment could be adapted to meet that demand.

However, the history of satellite telephony and broadband constellations shows that what is considered technically feasible is not necessarily what best meets the needs of operating companies or of potential users. Imagination, innovation and actual implementation are often very different things.

Elliptical orbits may not gain widespread use for communication, remaining a small niche in the communications niche that is satellite communication.

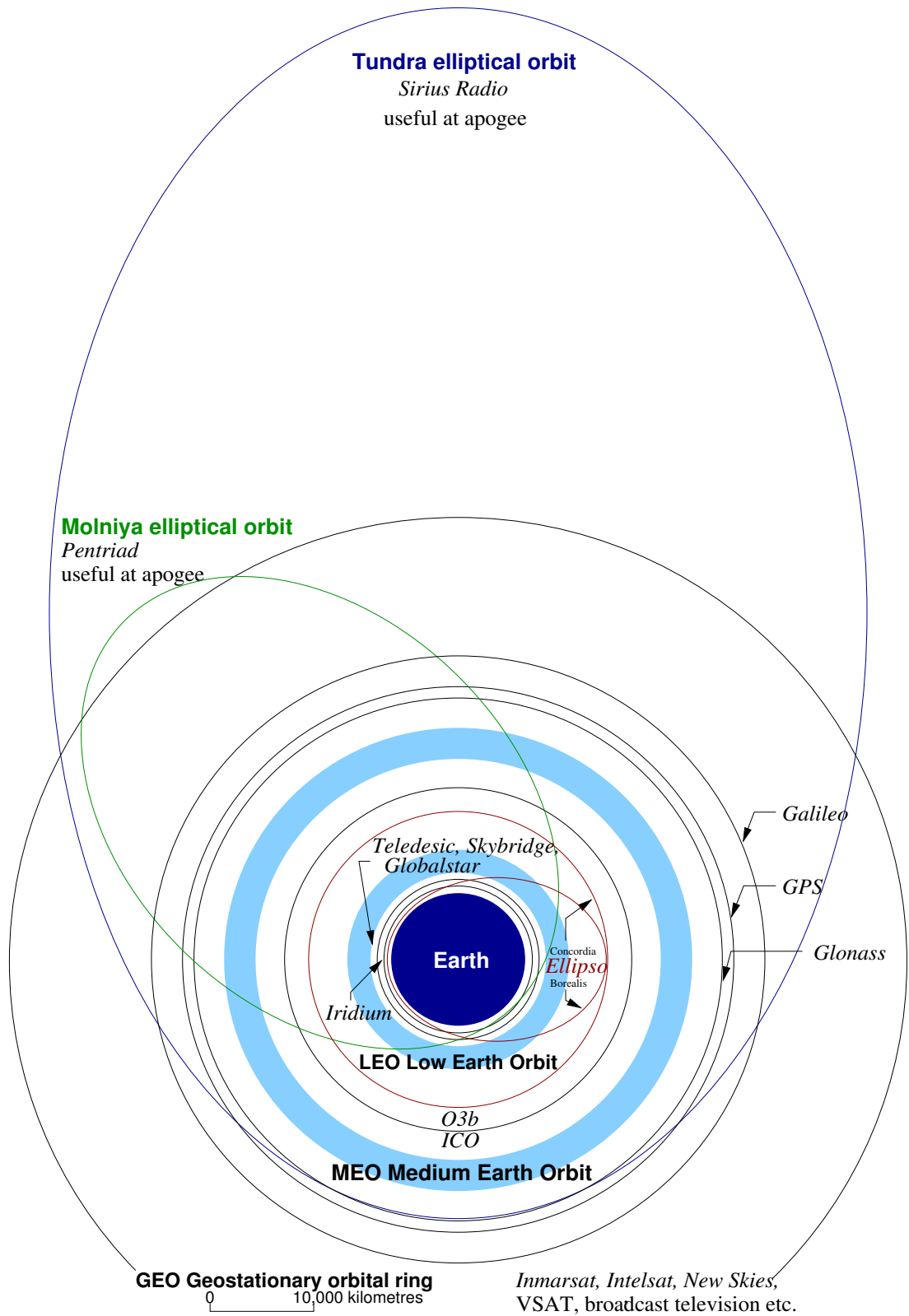

Fig. 1. **Orbital altitudes of satellite systems discussed**

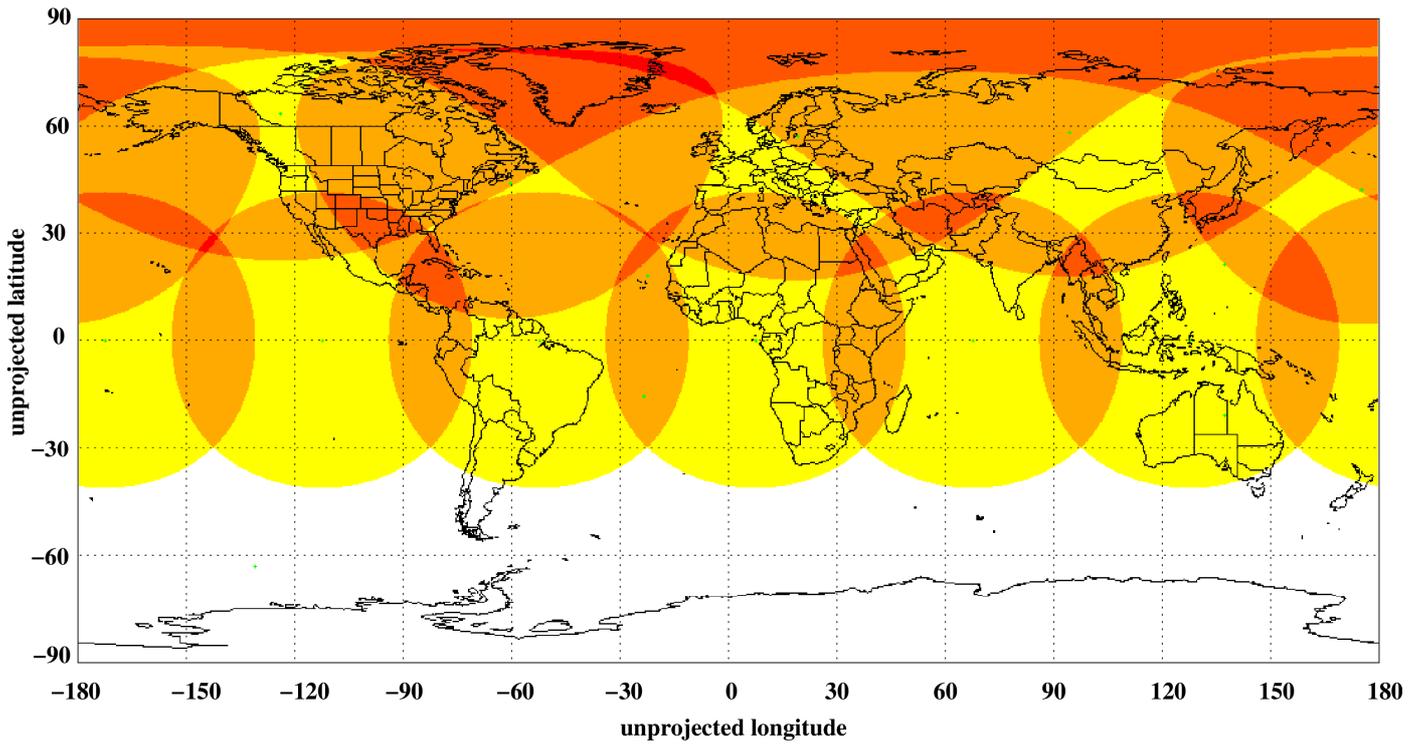

Fig. 2. *Ellipso* **constellation transmission, showing equatorial *Concordia* and elliptical *Borealis* subconstellations (*SaVi*)**

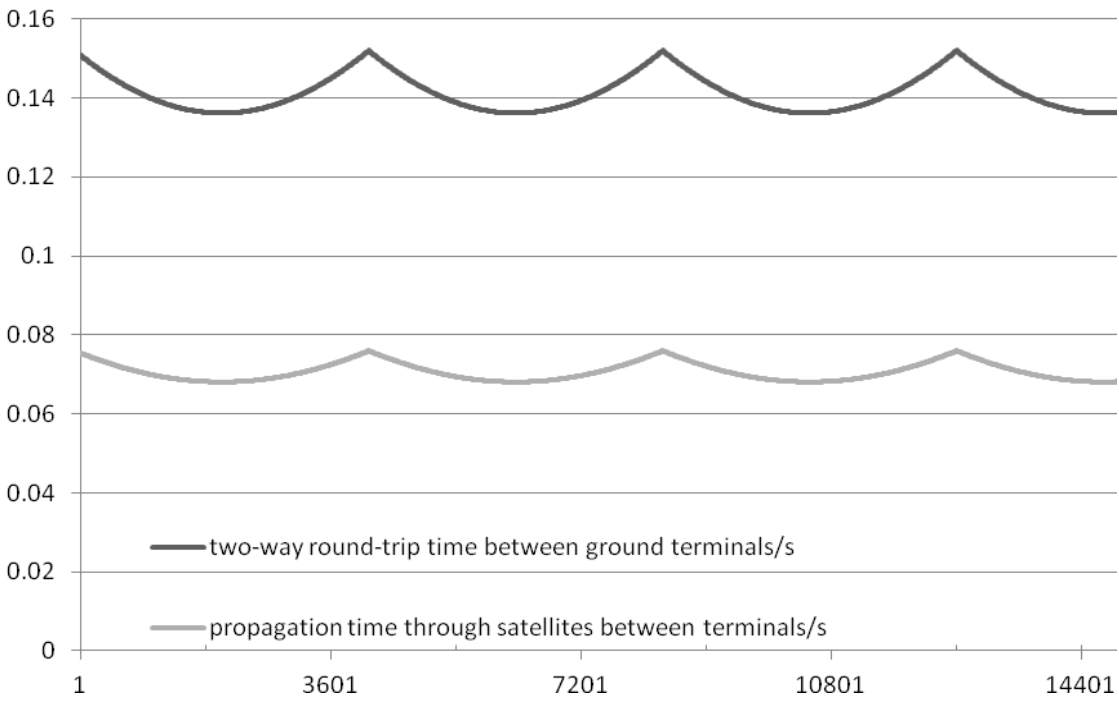

Communications between ground terminals through O3b satellites. Packet delay time/seconds *vs* time/seconds.

Fig. 3. **Sample *O3b* propagation delay, showing satellite handovers, simulated using *ns*-2**

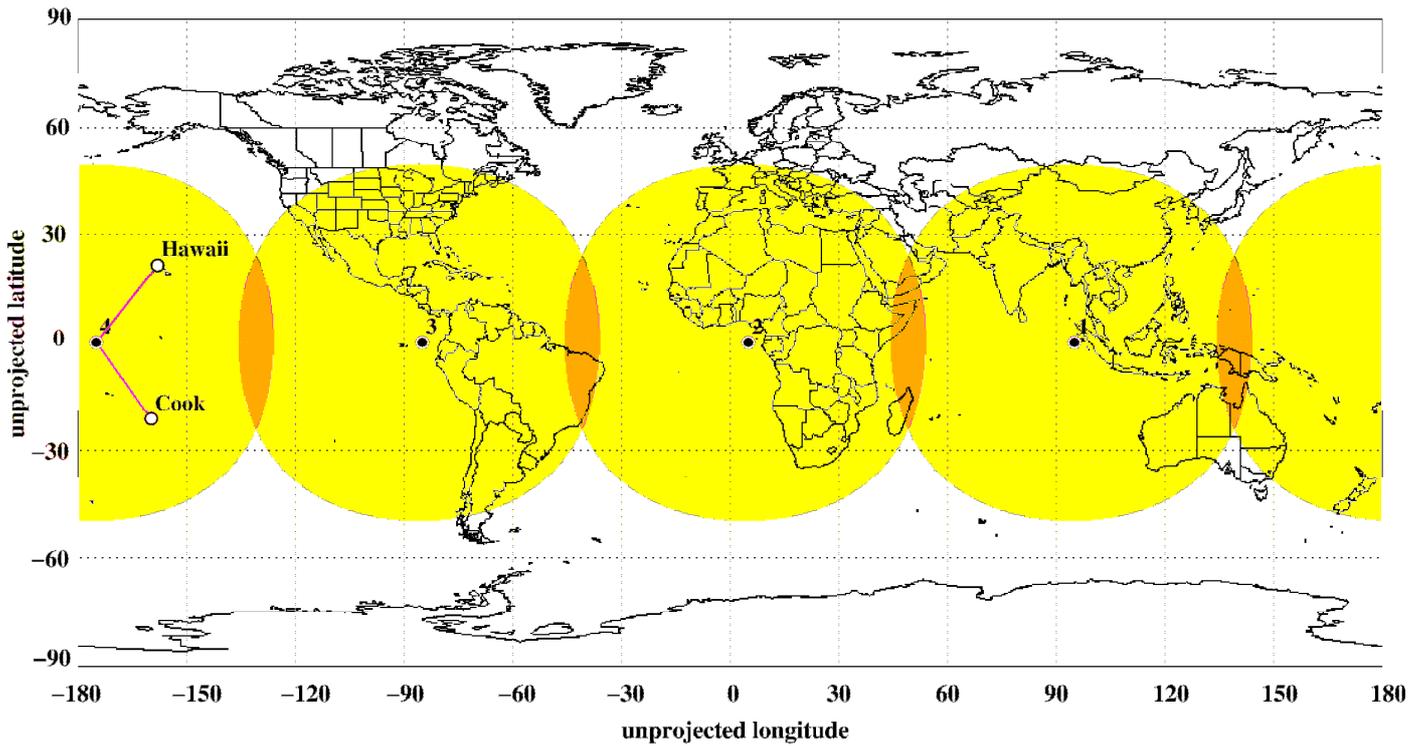

Example test communications between Gateway in Sunset Beach, Hawaii, and Avarua, Cook Islands.

Fig. 4. **Currently operational *O3b* constellation, showing coverage limits in testing with four active satellites (*SaVi*)**

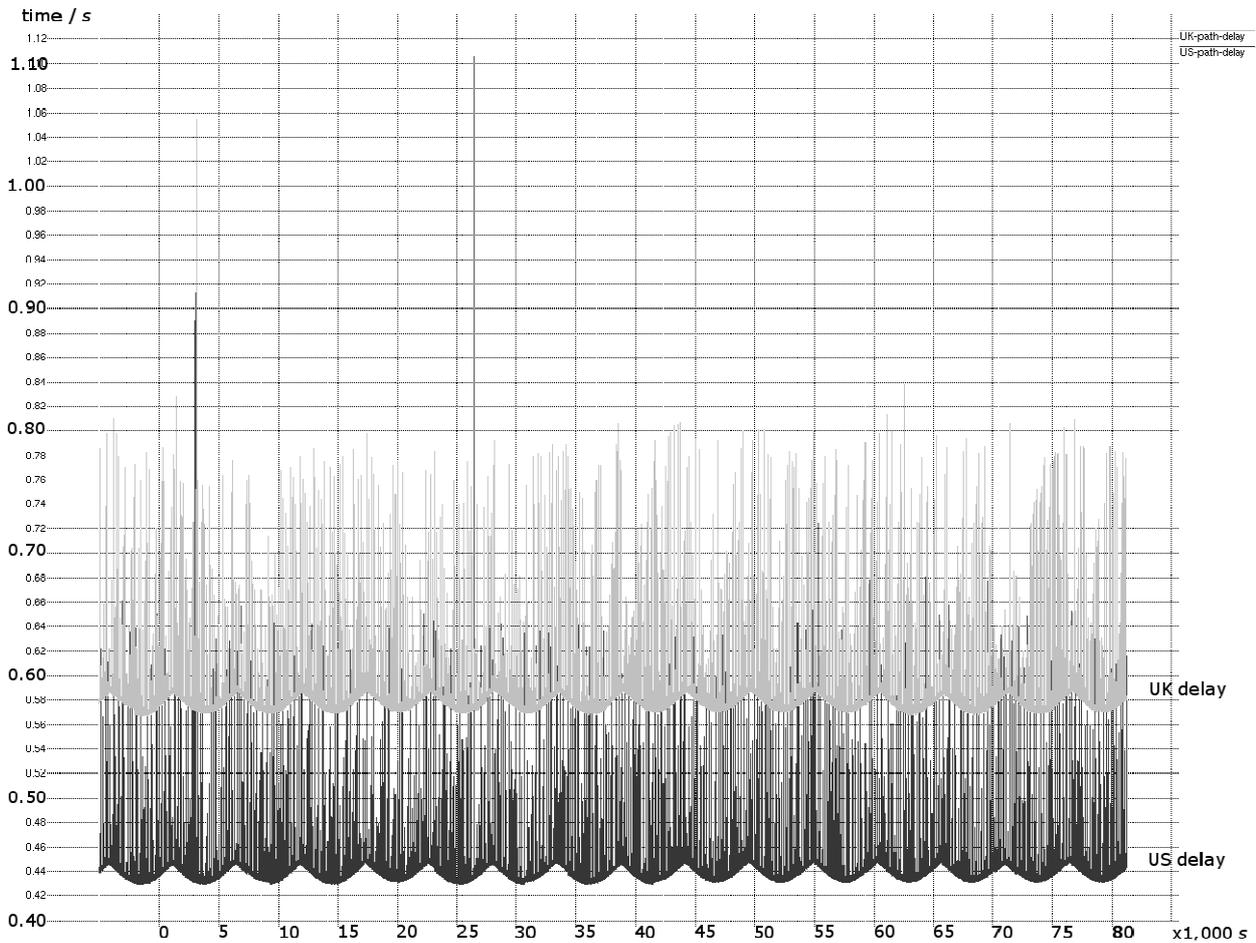

Fig. 5. **Sample internet path delays to Cook Islands via *O3b* over 24 hours, from computers in US and UK**

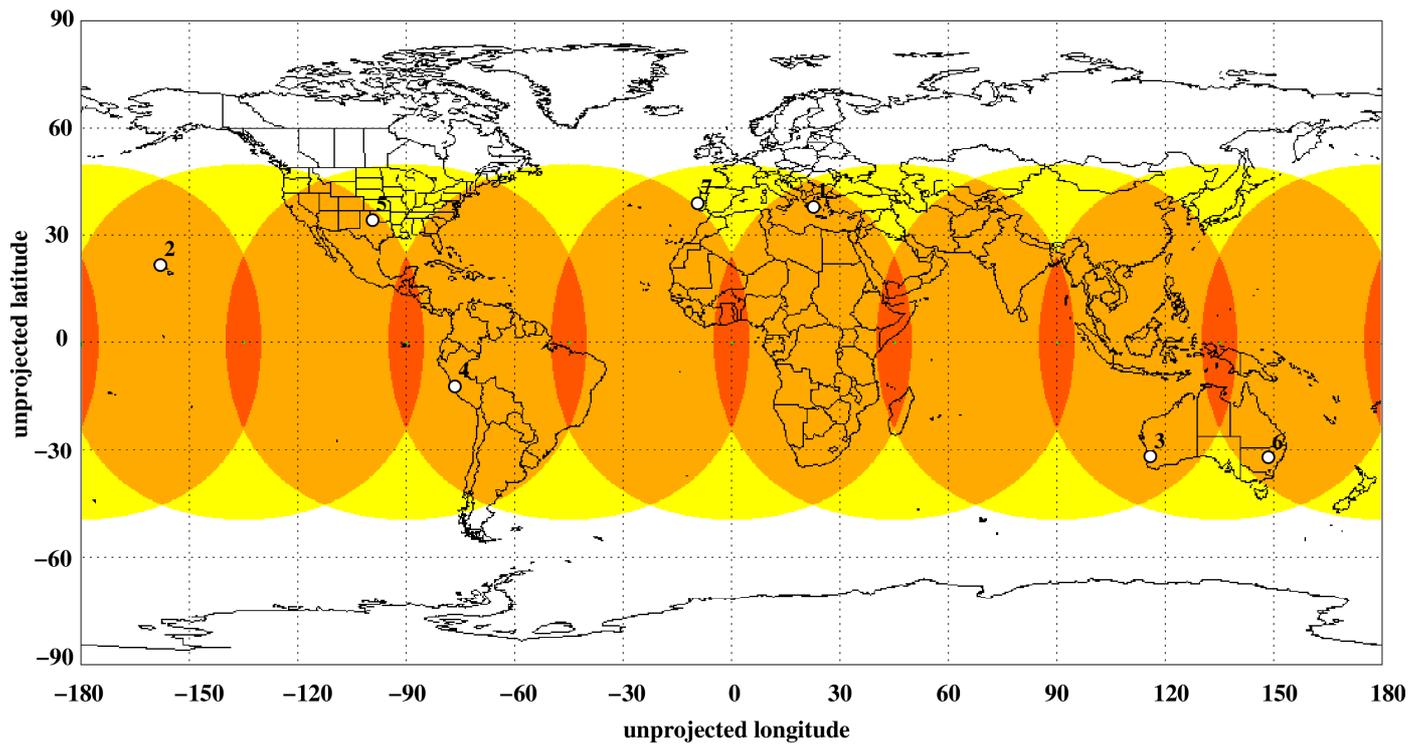

O3b gateway locations, in order of construction:
1. Nemea, Greece
2. Sunset Beach, Hawaii
3. Perth, Western Australia
4. Lurin, Peru
5. Vernon, Texas, United States
6. Dubbo, New South Wales, Australia
7. Sintra, Portugal

Fig. 6. **Possible future *O3b* constellation, showing seven constructed gateway stations and eight active satellites (*SaVi*)**

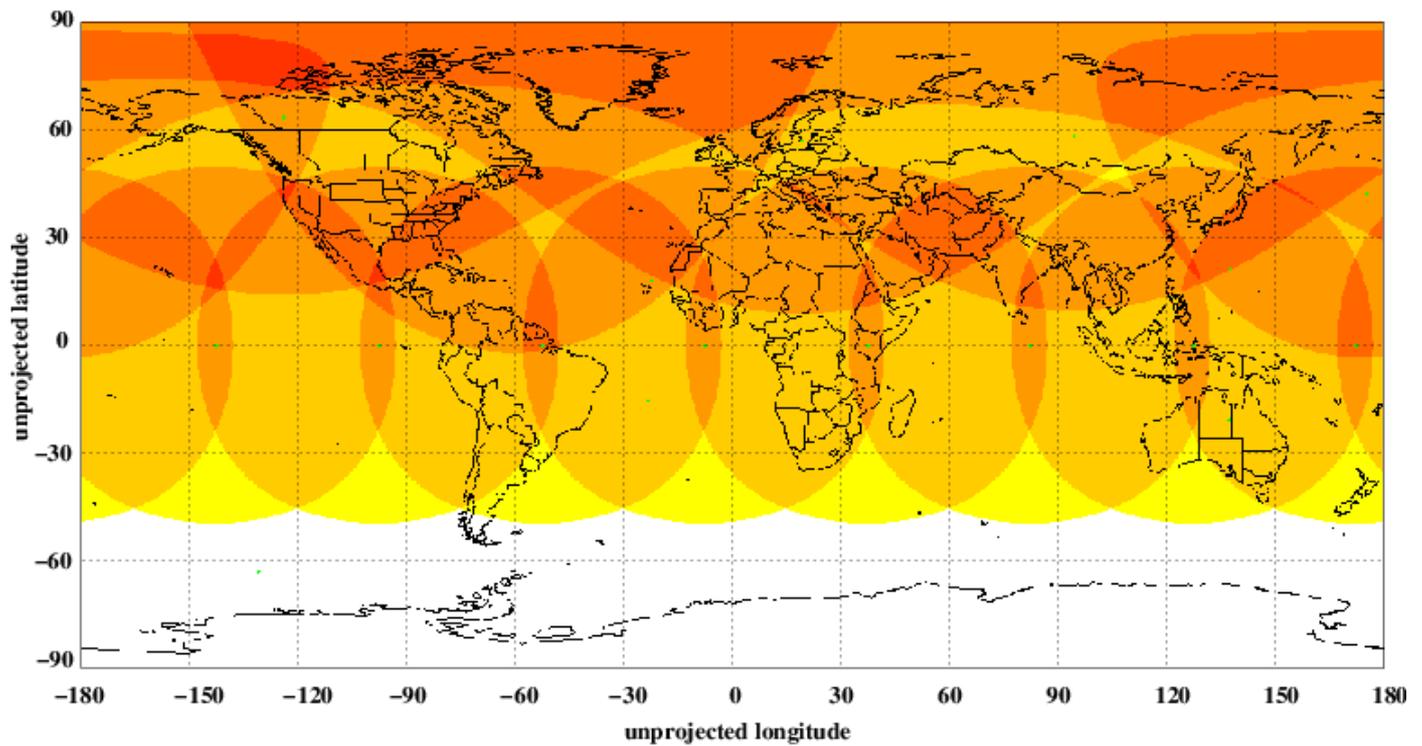

Fig. 7. **Coverage of enhanced *O3b* constellation, with eight active satellites and elliptical component (*SaVi*)**

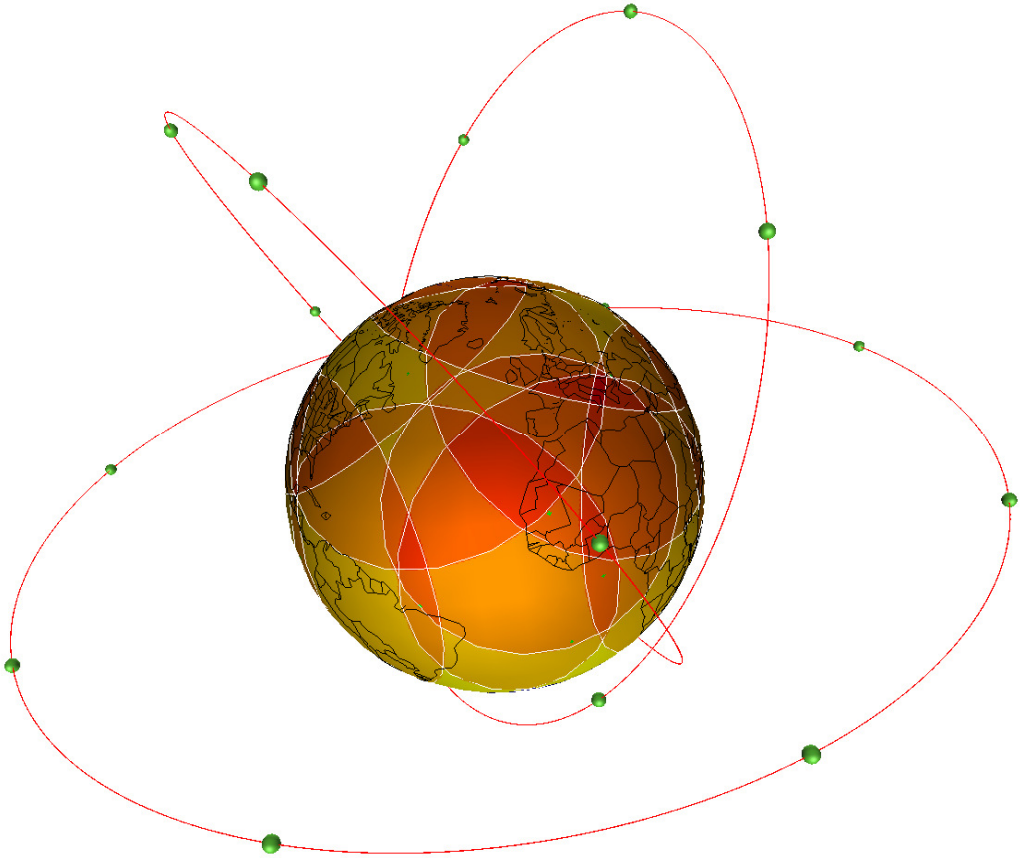

Fig. 8. **Enhanced *O3b* constellation, with eight MEO satellites and northern-hemisphere elliptical component (*SaVi*)**